\documentclass[prl,twocolumn,superscriptaddress,preprintnumbers,amsmath,amssymb,floatfix]{revtex4}  
\usepackage{graphicx}  
  
\newcommand{\tg}{t_{2g}}  
\newcommand{\eg}{e_g}

\begin{document}  
\title{Half-metallic ferromagnetism induced by dynamic electron  
  correlations in VAs}  
\author{L.~Chioncel}  
\affiliation{Institute of Theoretical Physics, Technical University of Graz,  
A-8010 Graz Austria}  
\affiliation{Faculty of Science, University of Oradea, RO-410087 Oradea, Romania}
\author{Ph.~Mavropoulos}  
\author{M.~Le\v{z}ai\'c}  
\author{S.~Bl\"ugel}  
\affiliation{Institut f\"ur Festk\"orperforschung, Forschungscentrum   
J\"ulich, D-52425 J\"ulich Germany}  
\author{E.~Arrigoni}  
\affiliation{Institute of Theoretical Physics, Technical University of Graz,   
A-8010 Graz Austria}  
\author{M.I.~Katsnelson}  
\affiliation{University of Nijmegen, NL-6525 ED Nijmegen, The Netherlands}  
\author{A.I.~Lichtenstein}  
\affiliation{Institute of Theoretical Physics, University of Hamburg, Germany}  
  
\begin{abstract}  
The electronic structure of the VAs compound in the zinc-blende
structure is investigated using a combined density-functional and
dynamical mean-field theory approach.  Contrary to predictions of a
ferromagnetic semiconducting ground state obtained by
density-functional calculations, dynamical correlations induce a
closing of the gap and produce a half-metallic ferromagnetic
state. These results emphasize the importance of dynamic correlations
in materials suitable for spintronics.
\end{abstract}

\maketitle

There is a growing interest in finding potential materials with high
spin polarization. Suitable candidates are Half-metallic ferromagnets
(HMF) which are metals for one spin direction and semiconductors for
the other \cite{deGroot83,ufn}. As a result, HMF are expected to show
a 100\% polarization and can in principle conduct a fully
spin-polarized current at low temperature. Therefore, HMF are ideal
candidates for potential applications in spintronics
\cite{ufn,sarma1}. 
These materials are 
equally interesting for basic research, in particular concerning the origin of
the half-metallic gap, the nature of inter-atomic exchange
interactions \cite{Kudrnovsky_NL}, its stability at elevated
temperatures \cite{LezaicTemp}, or the effect of electron correlations
\cite{ufn}.
  
Most theoretical efforts for understanding HMF are supported by  
first-principles calculations, based on density-functional theory  
(DFT). In fact the very discovery of HMF was due to such  
calculations~\cite{deGroot83}. DFT calculations are usually based on  
the Local Spin Density Approximation (LSDA) or the Generalized Gradient  
Approximation (GGA). These approximations have been proved very  
successful to interpret or even predict material properties in many  
cases, but they fail notably in the case of strongly-correlated  
electron systems. For such systems the so-called LSDA+$U$ (or GGA+$U$)  
method is used to describe static correlations, whereas dynamical  
correlations can be approached within the LSDA+DMFT (Dynamical  
Mean-Field Theory) \cite{anisDMFT,ourDMFT}. An important   
dynamical many-electron feature of half-metallic ferromagnets is the   
appearance of non-quasiparticle states \cite{ufn,edwards,IK} which can   
contribute essentially to the tunneling transport in heterostructures   
containing HMF \cite{ourtransport,falko}.  
  
Equally interesting materials for spintronics applications are
ferromagnetic semiconductors \cite{nagaev,ohno}.  Candidate
systems are ordered compounds such as europium chalcogenides (e.g.,
EuO) and chromium spinels (e.g., CdCr$_2$Se$_4$) \cite{nagaev}, as
well as diluted magnetic semiconductors (e.g., Ga$_{1-x}$Mn$_x$As)
\cite{ohno}. Unfortunately, all of them have Curie temperatures
much lower than room temperature. On the other hand, VAs in the
zinc-blende structure is, according to density-functional
calculations~\cite{Galanakis03}, a ferromagnetic semiconductor with a
high Curie temperature. Unlike CrAs~\cite{Akinaga00},
CrSb~\cite{Zhao01}, and MnAs~\cite{Okabayashi05}, VAs has not yet been
experimentally fabricated in the zinc-blende structure, but the
increasing experimental activity in the field of the (structurally
metastable) zinc-blende ferromagnetic compounds is promising in this
respect.
  
\begin{figure}[h]
\includegraphics[width=0.865\linewidth]{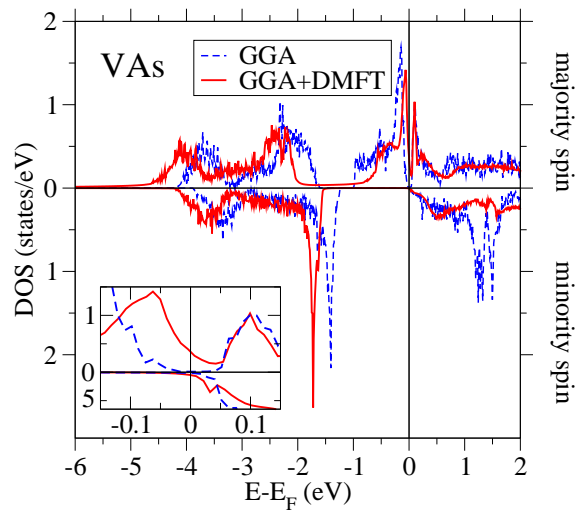}
\caption{(color online) DOS of VAs within the GGA (dashed blue line)
and GGA+DMFT (solid red line) for a temperature of $T=200K$, 
$U=2$~eV and 
$J=0.9$~eV.  Inset: Focus around $E_F$ showing the semiconducting gap within the GGA. To
ilustrate the minority-spin NQP states a ten times larger scale for the spin down channel
is used.
\label{200_vas_DOS}}
\end{figure}

In this Letter we investigate the effect of electronic interactions on  
VAs in the zinc-blende structure using dynamical mean-field theory. 
Our main result is displayed in Fig.~\ref{200_vas_DOS} (especially the inset):
While this material is expected to be a ferromagnetic semiconductor from 
density-functional theory (LSDA/GGA) or static LSDA+$U$ calculations, 
the inclusion of dynamic Coulomb correlations within the LSDA+DMFT approach predicts
a metallic behavior, due to the closure of the gap in the majority-spin band. 
Moreover, since the minority-spin band gap remains finite, the material is  found to be a  
half-metallic ferromagnet. To our knowledge, this is a first example  
in which dynamic correlations transform a semiconductor into a half
metal. This remarkable result demonstrates the relevance of many-body effects for spintronic  
materials.   
\begin{figure}[h]  
\includegraphics[angle=270,width=\linewidth]{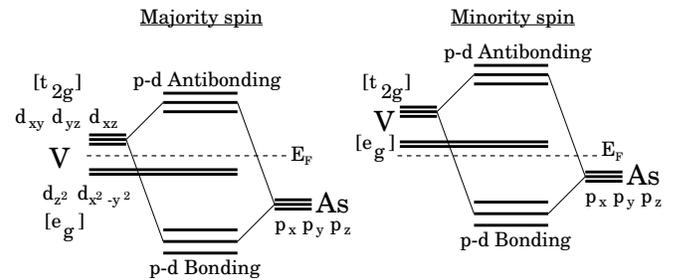}  
\caption{Schematic representation of the $p$-$d$ hybridization and  
bonding-antibonding splitting in VAs. \label{fig:bond_antibond}}  
\end{figure}  
  
The main features of the electronic structure
of VAs \cite{Galanakis03} are shown schematically in
Fig.~\ref{fig:bond_antibond}. The $\tg$ states hybridize with the
neighboring As $p$ states, forming wide bonding and antibonding hybrid
bands. In contrast, the $\eg$ states form mainly non-bonding and
narrow bands.  The Fermi level falls between the $\eg$ and the
antibonding $\tg$ in the majority-spin bands, and between the bonding
$\tg$ and the $\eg$ in the minority-spin bands.  
Thuse, the material is a ferromagnetic semiconductor, showing a very
narrow gap at $E_F$ (of the order of 50~meV) for majority-spin and a
much larger gap for minority spins. The spin moment, concentrated
mainly at the V atoms, is an integer of exactly $M=2\ \mu_B$ per unit
formula, which is obvious by counting the occupied bands of the two spin directions.
  
The exchange constants of VAs were calculated within GGA and adiabatic
spin dynamics. The energy $E(\vec{q})$ of frozen magnons is calculated
as a function of the wavevector $\vec{q}$ using the full-potential
linearized augmented plane wave method (FLAPW)~\cite{flapw} which
allows to evaluate the real-space exchange constants $J_{ij}$ by a
Fourier transform of $E(\vec{q})$. The procedure is similar to the one
used by Halilov {\it et al.}~\cite{exchange}. 
Using these exchange parameters in
a Monte Carlo simulation of the corresponding classical Heisenberg
Hamiltonian $E=-(1/2)\sum_{ij} J_{ij}\vec{M}_i\cdot\vec{M}_j$ (where
$\vec{M}_i$ and $\vec{M}_j$ are the magnetic moments at sites $i$ and
$j$), we obtain a Curie temperature $T_C=820$~K by the fourth-order
cumulant crossing point. This result agrees with 
the value of $T_C=830$~K calculated in Ref.~\cite{Sanyal03} using a
similar method.  The high Curie point is well above room temperature
making VAs a very promising candidate for applications in spintronics.
  
In order to take into account static correlations, we employ the
GGA+$U$ method (using the values $U=2$~eV, $J=0.9$~eV typical for 3$d$
transition metals \cite{today,arya06}). Accordingly, we used the GGA equilibrium 
lattice parameter,  $a=5.69\AA$, 
and a broadening $\delta$ of about $15K$,
which allows the majority spin gap to be clearly resolved. 
For different lattice parameters 
(eg, the InAs parameter) our LDA results agrees with the ones previously 
reported \cite{Sanyal03,Galanakis03}. The main difference between the GGA
DOS (see Fig.~\ref{200_vas_DOS}) and GGA+$U$ spectrum (not shown here)
is, as expected, that within the GGA+$U$ the occupied, localized
majority $\eg$ states are shifted to even lower energy, while the
unoccupied, minority $\eg$ states are shifted to higher energy.  The
semiconducting character does not change, since the $\eg$ and $\tg$
bands remain separated for both spins; the majority-spin gap slightly
increases but remains small.
  
In order to investigate dynamic correlation effects in VAs we used a
recently developed fully self-consistent in spin, charge and
self-energy LSDA+DMFT scheme \cite{EMTODMFT}.  This scheme uses for the 
LDA/GGA calculations the exact muffin-tin orbitals (EMTO) theory \cite{EMTO},
and the full charge density technique. This method combines the accuracy of 
the full-potential method and the efficiency of the muffin-tin potential 
method \cite{EMTO}. Correlation effects are treated in the framework of 
dynamical-mean-field theory with a spin-polarized T-matrix Fluctuation 
Exchange type of quantum impurity solver \cite{Katsnelson01,Leonid05}. The 
computational details are described in Refs.  \cite{Chioncel03,Chioncel05,EMTODMFT}.
 
In our calculations, we considered the standard representation of the
zinc-blende structure with an fcc unit cell containing four atoms: V
$(0,0,0)$, As $(1/4,1/4,1/4)$ and two vacant sites at
$(1/2,1/2,1/2)$ and $(3/4,3/4,3/4)$.  The charge density is calculated
by integrating the Green function along a complex contour 
up to the Fermi level. 
The multipole expansion of the charge density is cut off at
$l_{max}=8$, and the expansion of the Green function is cut off at
$l_{max}=3$. The convergence of the calculations was checked up to 
a number of 1505 k vectors and several sets between 22 and 30 complex 
energy points. The PBE~\cite{PW} parameterization 
of the GGA exchange correlation potential was used. The one-particle,
LSDA/GGA+DMFT Green function $G_{\sigma}(\vec{k};E)$ is related to the
LSDA/GGA Green function
$G_{\sigma}^{\mathrm{0}}(\vec{k};E)
 =\left[E+\mu-H_{\sigma}^{\mathrm{0}}(\vec{k})\right]^{-1}$
and to the local (on-site) self-energy $\Sigma_{\sigma}(E)$ via the
Dyson equation
\begin{equation}  
G_{\sigma}^{-1}(\vec{k};E) =  
E+\mu-H_{\sigma}^{\mathrm{0}}(\vec{k}) -  
\Sigma_{\sigma}(E) \label{eq:dyson}
\end{equation}  
$H_{\sigma}^{\mathrm{0}}(\vec{k})$ is the LSDA/GGA Hamiltonian,  
dependent on the Bloch vector $\vec{k}$, and the spin index  
$\sigma\in\{\uparrow,\downarrow\}$. $\mu$ is the chemical  
potential. The many-body effects beyond the LSDA/GGA are described by the  
multiorbital interacting Hamiltonian  
$\frac{1}{2}\sum_{{i \{m, \sigma \} }} U_{mm'm''m'''}  
c^{\dag}_{im\sigma}c^{\dag}_{im'\sigma'}c_{im'''\sigma'}c_{im''\sigma}$
where ${m}$ are local orbitals at site $i$; $c^{\dag}$ and $c$ denote
creation and destruction operators respectively. For the multiorbital
Hamiltonian the on-site Coulomb interactions are expressed in terms of
two parameters $U$ and $J$ \cite{ourDMFT}.  
In order to avoid ``double counting'',
the static part of the self-energy is subtracted, i. e. $\Sigma_{\sigma}(E)$ is
replaced with $\Sigma_{\sigma}(E)-\Sigma_{\sigma}(0)$, as 
$\Sigma_{\sigma}(0)$ is already included in 
the LSDA/GGA part of the Green's function. It has been proven that this type of 
``metallic'' double-counting is suitable for medium correlated d-electron systems
\cite{Petukhov03}.

The computational results for the GGA and GGA+DMFT densities of states are 
presented in Fig. \ref{200_vas_DOS}. The non-quasiparticle (NQP) states 
in the minority spin band are visible just above the Fermi 
level (inset), predicted also by previous calculations \cite{Chioncel03,Chioncel05}.
The weak spectral weight of NQP state is due to the 
fact that the Fermi level is close to the right edge of the minority spin gap, 
as discused for CrAs having a similar structure \cite{Chioncel05}. 
The origin of the  NQP states is connected to the ``spin-polaron'' processes: 
the spin-down low-energy electron excitations, forbidden for the HMF in the
one-particle picture, turn out to be possible as superpositions of
spin-up electron excitations and virtual magnons \cite{edwards,IK}.
The local spin moments at the V atoms do not change significantly
(less than 5\%).

\begin{figure}[h]  
\includegraphics[width=0.9\linewidth]{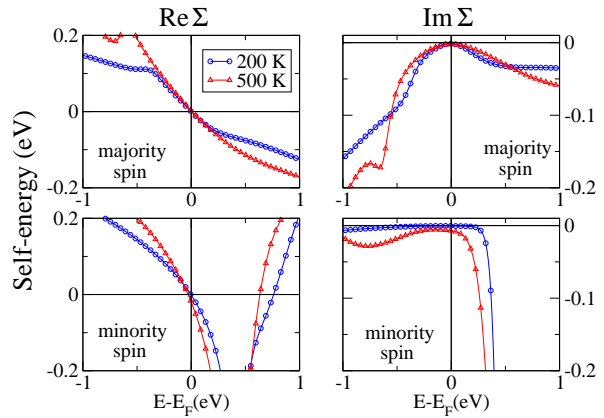} \\  
\caption{(color online) Energy dependence of real and imaginary parts of the  
self-energy $\Sigma_{\sigma}(E)$ for the $e_{g}$ orbitals.   
\label{sigma_eg}}  
\end{figure}

However, in the case of VAs another correlation effect appears: the
small majority-spin gap at $E_F$ closes, making the material half
metallic.  This prediction is the central result of our work.
\begin{figure}[h]  
\includegraphics[width=0.9\linewidth]{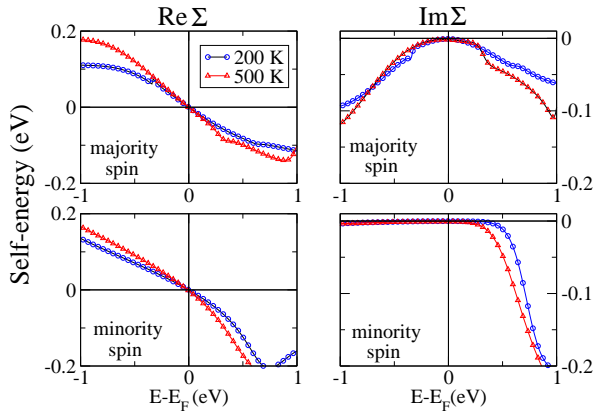} \\  
\caption{(color online) Energy dependence of real and imaginary parts of the  
self-energy $\Sigma_{\sigma}(E)$ for the $t_{2g}$ orbitals. \label{sigma_t2g}}  
\end{figure}  
In order to investigate the mechanism of the gap closure for the
majority spin channel, we look at the behavior of the electron
self-energy.  The quadratic form of the imaginary part of the
majority-spin self-energy, $\mathrm{Im}\Sigma_{\uparrow}(E)\sim
(E-E_F)^2$, visible in both Figs.~\ref{sigma_t2g} and \ref{sigma_eg},
indicates a Fermi liquid behavior, as opposed to
$\Sigma_{\downarrow}(E)$ which shows a suppression around $E_F$ due to
the band gap, as well as a peculiar behavior for $E>E_F$ related to
the existence of NQP states.  \cite{edwards,IK}.
From the Dyson equation~(\ref{eq:dyson}), one can see that the real
part of the self-energy, $\mathrm{Re}\Sigma_{\sigma}(E)$, causes a
shift of the LDA energy levels. Therefore, due to the non-zero
$\Sigma^{e_g}_{\uparrow}$, the $e_g$ orbitals in the close vicinity of
the Fermi level are pushed closer to $E_F$.  This renormalization of
the levels is connected to the large value of $\mathrm{Re}(\partial
\Sigma / \partial E)_{E_F} < 0$, pushing states in the vicinity of
$E_F$ more closer to $E_F$. This causes occupied levels to be
renormalized to higher energy and unoccupied levels to lower energy.
Notice that this effect is completely opposite to the GGA+U results,
discussed above.  In addition to this shifting, the $e_g$ peak is
broadened by correlations, its tail reaching over the Fermi level
(Fig.~\ref{200_vas_DOS}, inset).  Thus, our finite-temperature
GGA+DMFT calculations demonstrate the closure of the narrow gap in the
spin-up channel, which is produced by the correlation-induced
Fermi-liquid renormalization and spectral broadening.  At the same
time, NQP states appear for the minority spin channel just above
$E_F$.

The slope of the majority spin channel self-energy is almost a
constant as a function of temperature: $\mathrm{Re}(\partial
\Sigma_{\uparrow} / \partial E)_{E_F} \approx -0.4$ between 200~K and
500~K.  The physical content of this lies in the quasiparticle weight
$Z=(1-\partial \mathrm{Re} \Sigma_{\uparrow} / \partial E)^{-1}$,
measuring the overlap of the quasiparticle wave function with the
original one-electron wave function, having the same quantum numbers.
Our numerical results indicate that $Z \approx 0.7$ is quite
temperature independent for small temperatures.  Thus, we could expect
this renormalisation to hold down to zero temperature.  As a
consequence the closure of the gap in the majority channel is a
quantum effect, originating from the multiorbital nature of the local
Coulomb interaction (orbitals are squeezed towards $E_F$) rather than
an effect of temperature.  We have verified that a similar gap closure
is obtained for larger values of $U$, namely, $U=4$ and $6$~eV, 
although the latter values should be taken with some caution in our
FLEX calculation, which is in principle appropriate only in weak to
intermediate coupling. As a general tendency, increasing $U-J$ produce 
a stronger Fermi-liquid renormalization in the majority spin channel.
We have also verified that 
the same effect is evidenced for $J=0eV$.
 
Recent mean-field (LDA+U) calculations by Anisimov {\it et
al.}~\cite{Anisimov02} yield a first-order, semiconductor to
ferromagnetic metal transition as a function of doping in the
FeSi$_{1-x}$Ge$_x$ alloy.  In contrast, our calculation clearly shows
that in VAs the closure of the semiconducting gap for majority spins
cannot be captured by a static approach; dynamic correlation
contributions in the multiorbital model are required.

In summary, we have investigated the electronic structure and
correlation effects in the zinc-blende alloy VAs.  On the one hand,
our density-functional theory calculations within the GGA predicts
this material to be a ferromagnetic semiconductor with a tiny gap of
about 50~meV in the majority-spin DOS.  On the other hand, dynamical
effects described by LDA+DMFT destroy the narrow band gap and turn the
material into an half-metallic ferromagnet.  According to our results
the closure of the band gap is due to the multiorbital nature of the
local Coulomb interaction, and can be described as a strong
correlation-induced Fermi liquid-like renormalization of majority-spin
states accompanied by a lifetime broadening.  At the same time, in the
minority-spin channel non-quasiparticle states appear just above
$E_F$. We stress that these results are a consequence of the interaction 
of spin-up electrons with spin-flip excitations. Neither the closure of the gap nor the
non-quasiparticle states can be obtained in the static LDA+$U$
approximation.  Our LDA/GGA calculation supplemented by a Monte Carlo
simulation also predicts a high Curie temperature of 820~K, which
makes this material of interest for technological applications.  
We expect $T_c$ to be  not much affected by dynamical correlation, for the same
reason for which  the effective exchange interaction parameters
are not affected, as it was demonstrated in recent works \cite{Katsnelson01,Katsnelson00}.

The revealed half-metallic (instead of semiconducting) behavior has
important consequences in the potential applications of VAs in
spintronics. In contrast to all-semiconductor-based spin-injection
devices~\cite{sarma1} which avoid the resistivity mismatch problem,
half-metals can be applied in giant magnetoresistance or, if interface
states are eliminated~\cite{Mavropoulos05}, in tunneling
magnetoresistance. As our calculations shows, in the prediction of new
spintronic materials, correlation effects play a decisive role.  While
in some materials these are detrimental for half-metallicity due to
the introduction of spectrum in the minority spin gap
\cite{Chioncel03}, the present case is an example in which
correlations turn out to be favorable for a high spin polarization.
The metallic nature of the majority spin channel would be visible in
resistivity measurements.  Therefore, the experimental realization of
zinc-blende VAs would provide a test of our prediction. Further research
should address the issue of the stability of the half-metallic 
ferromagnetic state in a zinc blende structure. Some work 
in this direction has been already carried out \cite{Shirai01,Xie03}. 
 DMFT total-energy calculations along the lines described in Ref.
\cite{zunger} are in progress. 

We thank Dr.~R.A.~de Groot for helpful discussions. 
We acknowledge financial support by the Research Center J\"ulich (LC) and by the 
Austrian science fund FWF, project No. P18505-N16 (LC and EA).

\end{document}